\documentclass[%
pra,%
preprint,%
onecolumn,%
titlepage,%
nofootinbib,%
amsmath,%
]{revtex4}%

\begin{document}
\title{Electron shielding of the nuclear magnetic moment\\ in hydrogen-like atom}
\author{V.~G.~Ivanov}
\email{ivanov.vg@gao.spb.ru} \affiliation{Pulkovo Observatory, St.Petersburg, 196140,
Russia} \affiliation{D.~I. Mendeleev Institute for Metrology, St.Petersburg, 190005, Russia}
\author{S.~G.~Karshenboim}
\email{s.g.karshenboim@vniim.ru} \affiliation{D.~I. Mendeleev Institute for Metrology,
St.Petersburg, 190005, Russia} \affiliation{Max-Planck-Institut f\"ur Quantenoptik, Garching,
85748, Germany}
\author{R.~N.~Lee}
\email{r.n.lee@inp.nsk.su} \affiliation{Budker Institute of Nuclear Physics, 630090,
Novosibirsk, Russia}

\begin{abstract}
The correction to the wave function of the ground state in a
hydrogen-like atom due to an external homogenous magnetic field is
found exactly in the parameter $Z\alpha$. The $j=1/2$ projection of
the correction to the wave function of the $ns_{1/2}$ state due to
the external homogeneous magnetic field is found for arbitrary $n$.
The $j=3/2$ projection of the correction to the wave function of the
$ns_{1/2}$ state due to the nuclear magnetic moment is also found
for arbitrary $n$. Using these results, we have calculated the
shielding of the nuclear magnetic moment by the $ns_{1/2}$ electron.
\end{abstract}

\maketitle

\section{Introduction and Conclusions}

Nowadays, the measurements of magnetic properties of a bound
electron or nucleus in simple atoms have reached an impressive level
of precision. This progress stimulates theoretical investigations of
subtle effects which contribute to the magnetic moment of a
hydrogen-like atom. To some extent, a nucleus can be considered as a
source of the Coulomb field distorted at small distances due to the
nuclear size effect. However, the present experimental accuracy
leads to the necessity to go beyond this approximation. For a
spinless nucleus one has to take into account the recoil effects,
which lead to the corrections in $m_e/m_N$ ($m_e$ is the mass of
electron and $m_N$ is the nuclear mass). For the nucleus with a
nonzero spin and magnetic moment, one also has to take into account
the interaction of the nuclear magnetic moment with the electron and
external magnetic field $\boldsymbol{B}$. E.g., for a spin-$1/2$
nucleus, these effects can be described by the Hamiltonian
\cite{bethe}
\begin{equation}\label{eq:ham_mag}
{\cal H}_{nlj} = \frac{e}{2m_e} \,g_e^\prime\,\boldsymbol{j}_e
\cdot\boldsymbol{B} - \frac{e}{2m_p}\,
g_N^\prime\,\boldsymbol{S}_N\cdot\boldsymbol{B} + A\;
\boldsymbol{j}_e \cdot\boldsymbol{S}_N + {\rm higher~order~terms}\;,
\end{equation}
which is defined in the space of spin and angular moment variables
for the hyperfine components of the $nl_j$ state in the
hydrogen-like atom. Here, $e$ is the elementary charge (positive),
$m_{e/p}$ is the electron/proton mass, $\boldsymbol{j}_e$ is the
electron total angular momentum, $\boldsymbol{S}_N$ is the nuclear
spin and $A$ is the hyperfine-interaction parameter. The
relativistic units in which $\hbar=c=1$ are applied through the
paper and $e^{2}=\alpha\approx1/137$ is the fine structure constant.
The bound $g$ factors of the electron and nucleus ($g^\prime_{e/N}$)
are somewhat different from their free values $g^{(0)}_{e/N}$ due to
bound effects which increase with $Z$.

Study of such a bound correction for the nuclear $g$ factor (i.e.
the screening or shielding effect) for certain states was done some
time ago in \cite{moo,pyp}. This result was checked for the $1s$
state in \cite{cjp02}. An important detail of the latter calculation
is the related correction to the wave function which was partly
obtained for the $1s$ state in \cite{jetp00,jetp01} for other
purposes. Actually, one of the purposes of \cite{cjp02} was to
verify the previously obtained \cite{jetp00,jetp01} results for the
wave function.

In this paper, we consider the effect of shielding of the nuclear magnetic moment by the electron in $ns_{1/2}$ state, which results in the
difference between $g_N^\prime$ and $g_N^{(0)}$. For this purpose, we calculate the corrections to the electron wave function caused by
both an external magnetic field and the nuclear magnetic moment. The operators corresponding to these effects, have vector character and
the correction to the wave function of an $s$ state is a mixture of $j=1/2$ and $j=3/2$ parts. The form of these parts is quite different. In
particular, the $j=1/2$ projection of the correction due to the external magnetic field for the $ns$ state  has the form of some polynomial
operator acting on the unperturbed wave function, Eq.~(\ref{eq:deltaPsi12}). The $j=3/2$ projection for this correction can not be
represented in the similar form. For the specific case of $1s$ state, this correction is calculated below. For the correction due to the nuclear
magnetic moment the situation is the opposite: the $j=3/2$ projection has the form of some polynomial operator acting on the unperturbed
wave function (see~Eq.~(\ref{eq:deltaPsi32}) , while the $j=1/2$ projection can not be expressed in such a simple form (see Appendix).

Using the derived corrections, we have calculated the shielding of
the nuclear magnetic moment in $ns_{1/2}$ state. The expansion of
our result (see Eq.~(\ref{eq:kns}) for a result exact in $Z\alpha$)
is
\[
g_N^\prime=g_N^{(0)}\times\left\{1-\frac{\alpha(Z\alpha)}{3n^{2}}\left[
1+\frac{132n-35}{36n^{2}}(Z\alpha
)^{2}+\frac{370n^{3}+342n^{2}-492n+69}{72n^{4}}(Z\alpha)^{4}\right.\right.
\]%
\[
\left.\left. +\frac{13208n^{5}+15048n^{4}-8552n^{3}-25320n^{2}+17064n-1641}%
{1728n^{6}}(Z\alpha)^{6}+\dots \right]\right\}\,.
\]
For the specific case of the ground state, this expression agrees with those in \cite{moo,pyp}.
Our method of calculations is, in part, similar to theirs and the related results on the wave
function in our paper and in \cite{moo,pyp} are also consistent.

The paper is organized as follows. In Section~\ref{sec:ii} we
calculate the correction to the wave function of the ground state in
a hydrogen-like atom due to an external constant magnetic field. We
check this correction by calculating the magnetic polarizability and
the shielding of the nuclear magnetic moment in the ground state,
which are known. In Section~\ref{sec:iii} we calculate the shielding
of the nuclear magnetic moment by the $ns_{1/2}$ electron.

\section{Corrections to the wave function of the ground state in the external magnetic
field}
\label{sec:ii}

In the present Section we calculate the correction to the wave
function of the ground state due to an external constant magnetic
field ${\bf B}$.

The unperturbed wave functions of the ground state with $j_{z}=\pm1/2$ have the form
\begin{align*}
\psi_{\pm} & =C\,\left( mr\right)^{\gamma-1}\exp\left( -Z\alpha mr\right) \left(
\begin{array}
[c]{c}%
\sqrt{1+\gamma}\;\varphi_{\pm}\\
\sqrt{1-\gamma}\;i\left( \boldsymbol{\sigma}\cdot\boldsymbol{n}\right)
\varphi_{\pm}%
\end{array}
\right)\,,
\end{align*}
where
\[
\boldsymbol{n} =\boldsymbol{r}/r\,,\quad
C =\frac{(2Z\alpha)^{\gamma+1/2}m^{3/2}}{\sqrt{8\pi\Gamma(2\gamma+1)}}\,,
\]
\[
\varphi_{+}=\left(
\begin{array}
[c]{c}%
1\\
0
\end{array}
\right)\,,\quad\varphi_{-}=\left(
\begin{array}
[c]{c}%
0\\1
\end{array}
\right)\,,\quad \gamma=\sqrt{1-\left( Z\alpha\right)^{2}}\,.
\]

Let the external magnetic field be directed along $z$-axis. The
Hamiltonian has the form%
\begin{align*}
H & =\boldsymbol{\alpha}\cdot(\boldsymbol{p}-e\boldsymbol{A}%
)-Z\alpha/r+\gamma^{0}m=H_0-e(\boldsymbol{\alpha}\cdot\boldsymbol{A})\,,\\
\boldsymbol{A} & =\frac{1}{2}\;\boldsymbol{B}\times\boldsymbol{r}\,.
\end{align*}

The states $\psi_{\pm}$ form a diagonal basis for the
perturbation. The correction to the wave function $\psi_{+}$ is
expressed as
\[
\delta\psi=-eG_{r}\left( \boldsymbol{\alpha}\cdot\boldsymbol{A}\right)
\psi_{+}=\frac{eB}{2}G_{r}\left( \boldsymbol{\alpha}\times\boldsymbol{r}%
\right)_{z}\psi_{+}\,,
\]
where $G_{r}$ is the reduced Green function determined as
\begin{align}
G_{r}=\left(1-\sum_{\pm}\left| \psi_{\pm}\right\rangle \left\langle \psi_{\pm}\right|\right)
\frac{1}{\varepsilon-H_0+i0}\left(1-\sum_{\pm}\left| \psi_{\pm}\right\rangle \left\langle \psi_{\pm}\right|\right)
\end{align}
and $\varepsilon=m\gamma$ is the energy of the ground state. In order to
determine $\delta \psi$ explicitly, we first separate parts with
$j=1/2$ and $j=3/2$. They can be expressed in terms of the four
functions $g_{1},\,h_{1},\,g_{2},\,h_{2}$ yet to be determined:
\begin{align}
\delta\psi_{B} & =\delta\psi_{B}^{1/2}+\delta\psi_{B}^{3/2}\,,\nonumber\\
\delta\psi_{B}^{1/2} & =C\,\frac{eB}{2m}\,\left( mr\right)^{\gamma-1}%
\exp\left( -Z\alpha mr\right) \left(
\begin{array}
[c]{c}%
\sqrt{1+\gamma}\;g_{1}\left( r\right) \varphi_{+}\\
\sqrt{1-\gamma}\;h_{1}\left( r\right) i\left( \boldsymbol{\sigma}%
\cdot\boldsymbol{n}\right) \varphi_{+}%
\end{array}
\right)\,, \label{eq:dpsi1(1/2)}\\
\delta\psi_{B}^{3/2} & =C\,\frac{eB}{2}\,\left( mr\right)^{\gamma-2}%
\exp\left( -Z\alpha mr\right) \left(
\begin{array}
[c]{c}%
\sqrt{1+\gamma}\;g_{2}(r)\;\left[ z\left( \boldsymbol{\sigma}\cdot
\boldsymbol{n}\right) -r/3\right] \varphi_{+}\\
\sqrt{1-\gamma}\;h_{2}(r) \; i\left[ z-\left( \boldsymbol{\sigma}\cdot
\boldsymbol{r}\right)/3\right] \varphi_{+}%
\end{array}
\right)\,. \label{eq:dpsi1(3/2)}%
\end{align}

Using the fact that $H_0$ commutes with $\boldsymbol{j}$ and that
the projection operators on the subspaces with $j=1/2$ and $j=3/2$
in our case can be represented as
\begin{equation}
P_{1/2}=\frac{5}{4}-\frac{\boldsymbol{j}^{2}}{3}\,,\quad P_{3/2}%
=\frac{\boldsymbol{j}^{2}}{3}-\frac{1}{4}\,, \label{eq:Projectors}%
\end{equation}
we have%
\begin{align}
\delta\psi_{B}^{1/2} & =-C\,\frac{eB}{3m}G_{r}\,\left( mr\right)^{\gamma
}\exp\left( -Z\alpha mr\right) \left(
\begin{array}
[c]{c}%
\sqrt{1-\gamma}\varphi_{+}\\
\sqrt{1+\gamma}i\left( \boldsymbol{\sigma}\cdot\boldsymbol{n}\right)
\varphi_{+}%
\end{array}
\right)\,, \label{eq:dpsi2(1/2)}\\
\delta\psi_{B}^{3/2} & =C\,\frac{eB}{2}\,G_{r}\,\left( mr\right)
^{\gamma-1}\exp\left( -Z\alpha mr\right) \left(
\begin{array}
[c]{c}%
\sqrt{1-\gamma}\left[ z\left( \boldsymbol{\sigma}\cdot\boldsymbol{n}\right)
-r/3\right] \varphi_{+}\\
\sqrt{1+\gamma}i\left[ z-\left( \boldsymbol{\sigma}\cdot\boldsymbol{r}%
\right) /3\right] \varphi_{+}%
\end{array}
\right)\,. \label{eq:dpsi2(3/2)}%
\end{align}

Acting by the operator $\varepsilon-H_0$ on both sides of
Eqs.~(\ref{eq:dpsi2(1/2)}, \ref{eq:dpsi2(3/2)}) and using
Eqs.~(\ref{eq:dpsi1(1/2)}, \ref{eq:dpsi1(3/2)}), we obtain two
systems of equations corresponding to $j=1/2$ and $j=3/2$.

\subsection{The $j=1/2$ projection of the correction to the wave function of the ground state}

Taking into account that
\[
\left( \varepsilon -\mathrm{H_{0}}\right) G_{r}=1-\left|
\psi_{+}\right\rangle \left\langle \psi_{+}\right| -\left|
\psi_{-}\right\rangle \left\langle \psi_{-}\right|\;,
\]
we
obtain the following system of equations for the contribution of $j=1/2$:%
\begin{align}
g_{1}^{\prime} & =\left( \frac{1-\gamma}{r}+mZ\alpha\right) \left(
g_{1}-h_{1}\right) +\frac{Z\alpha\left( 2\gamma+1\right) }{3\left(
1+\gamma\right) }-\frac{2mr}{3}\,,\nonumber\\
h_{1}^{\prime} & =\left( \frac{1+\gamma}{r}-mZ\alpha\right) \left(
g_{1}-h_{1}\right)-\frac{Z\alpha\left( 2\gamma+1\right)
}{3\left( 1-\gamma\right) }+\frac{2mr}{3}\,. \label{eq:gh1equation}%
\end{align}

The boundary conditions can be easily determined from the
representation of the reduced Green function in terms of the
eigenfunctions of the Hamiltonian. Since the wave functions of the
$ns_{1/2}$ states behave as $r^{\gamma-1}$ at the origin, we
conclude that the functions $g_{1}$ and $h_{1}$ should be regular at
this point.

Let us pass to the functions $g_{1\pm}=g_{1}\pm h_{1}$. For $g_{1-}$
we obtain a closed equation:
\[
g_{1-}^{\prime}+2\left( \frac{\gamma}{r}-mZ\alpha\right) g_{1-}%
=\frac{2}{3}\frac{2\gamma+1}{Z\alpha}-\frac{4mr}{3}\,.%
\]

The general solution of this equation is
\[
g_{1-}=\frac{2r}{3Z\alpha}+C_{1}(m r)^{-2\gamma}\exp\left( 2Z\alpha mr\right)\,.
\]

Due to the boundary conditions, $C_{1}=0$. Using
Eq.~(\ref{eq:gh1equation}), we obtain
\begin{align*}
g_{1} & = \frac{Z\alpha\left( 2\gamma+3\right) }{3\left( 1+\gamma\right)
}r+C_{2}\,,\\
h_{1} & = -\frac{Z\alpha\left( 2\gamma-1\right) }{3\left( 1-\gamma
\right) }r+C_{2}\,.%
\end{align*}

The constant $C_{2}$ is completely determined by the condition of
the orthogonality of $\delta\psi^{1/2}$ and $\psi_{+}$. Finally, we
obtain
\begin{align*}
g_{1} & =\frac{Z\alpha\left( 2\gamma+3\right) }{3\left( 1+\gamma\right)
}r-\frac{2\gamma+1}{3m}\;,\\
h_{1} & =-\frac{Z\alpha\left( 2\gamma-1\right) }{3\left(
1-\gamma\right) }r-\frac{2\gamma+1}{3m}\;,
\end{align*}
which agrees with \cite{pyp,jetp01}.

\subsection{The $j=3/2$ projection of the correction to the wave function of the ground state}

We obtain the following system of equations for the contribution of
$j=3/2$:
\begin{align*}
g_{2}^{\prime} & =-\left( \frac{2+\gamma}{r}-mZ\alpha\right) g_{2}-\left(
\frac{1-\gamma}{r}+mZ\alpha\right) h_{2}+mr\,,\\
h_{2}^{\prime} & =\left( \frac{1+\gamma}{r}-mZ\alpha\right) g_{2}+\left(
\frac{2-\gamma}{r}+mZ\alpha\right) h_{2}-mr\,.
\end{align*}

Again, moving to $g_{2\pm}=g_{2}\pm h_{2}$, we have
\begin{gather}
g_{2-}=-rg_{2+}^{\prime}\label{eq:f2-}\,,\\
r\,g_{2+}^{\prime\prime}+\left( 2\gamma+1-2Z\alpha mr\right) g_{2+}^{\prime
}-\frac{3}{r}\,g_{2+}=-2mr\,.\nonumber
\end{gather}

The first equation determines the function $g_{2-}$ via $g_{2+}$.
The general solution of the second equation is
\begin{align*}
g_{2+} & =C_{1}(mr)^{-\nu-\gamma}\,{}_{1}F_{1}\left( -\nu-\gamma;\;1-2\nu;\;2Z\alpha
r\right) +C_{2}\left( mr\right)^{\nu-\gamma}\,{}_{1}F_{1}\left( \nu
-\gamma;\;1+2\nu;\;2Z\alpha mr\right) \\
& -\frac{2\left( mr\right)^{2}}{4\gamma+1}\,{}_{2}F_{2}\left(
2,1;\;3+\gamma+\nu,3+\gamma-\nu;\;2Z\alpha mr\right)\,,
\end{align*}
where $\nu=\sqrt{4-\left( Z\alpha\right)^{2}}$.

Again, $C_{1}=0$ due to the boundary conditions at the origin. The
constant $C_{2}$ is fixed to suppress the exponential growth of
$g_{2+}$ as $r\to \infty$. In order to determine it, we use the
following integral representation of $_{2}F_{2}$ (see, e.g.
 \cite{3f2slat}):
\begin{align*}
\frac{_{2}F_{2}\left( 2,1;a,b;x\right) }{\left( a-1\right) \left(
a-2\right) \left( b-1\right) } & = \int_{0}^{1}dz\, z\left( 1-z\right)
^{a-3}\int_{0}^{1}dt\, t^{b-2}\exp\left[ z\left( 1-t\right) x\right]
\,.
\end{align*}

We rewrite the integral over $t$ as
$\int_{0}^{1}dt=\int_{0}^{\infty} dt-\int_{1}^{\infty}dt$. The first
integral results in the $\Gamma$-function, and we obtain the
integral representation for the $_{1}F_{1}$ function. Making the
substitution $t\rightarrow1/t$ in the second term, we arrive at
\begin{align*}
_{2}F_{2}\left( 2,1;\;a,b;\;x\right) & =x^{1-b}\frac{\Gamma\left( a\right)
\Gamma\left( b\right) \Gamma\left( 3-b\right) }{\Gamma\left(
1+a-b\right) } {}_{1}F_{1}\left( 3-b;\;1+a-b;\;x\right) \\
& -\left( a-1\right) \left( a-2\right) \left( b-1\right) \int_{0}%
^{1}dz\,z\left( 1-z\right)^{a-3}\int_{0}^{1}dt\,t^{-b}\exp\left[ z\left(
1-1/t\right) x\right]
\end{align*}

In our case, $a=3+\gamma+\nu$, $b=3+\gamma-\nu$, $x=2Z\alpha mr$ and
the first term contains $_{1}F_{1}\left( \nu-\gamma;1+2\nu;2Z\alpha
mr\right) $. Note that the second term in the right-hand side does
not grow exponentially as $r\rightarrow\infty$. Thus, we have
\[
C_{2}=\frac{2\,\Gamma\left( 3+\gamma+\nu\right) \Gamma\left( 3+\gamma
-\nu\right) \Gamma\left( \nu-\gamma\right) }{m\left( 1+4\gamma\right)
\Gamma\left( 1+2\nu\right) }\left( 2Z\alpha\right)^{\nu-\gamma-2},
\]
and the following integral representation for $g_{2+}$ is valid:%
\[
g_{2+}=2mr^{2}\left( 1+\gamma+\nu\right) \int_{0}^{1}dz\,z\left(
1-z\right)^{\gamma+\nu}\int_{0}^{1}dt\,t^{-\gamma+\nu-3}\exp\left[ z\left(
1-1/t\right) 2Z\alpha mr\right]
\,.
\]

The function $g_{2-}$ is determined by Eq.~(\ref{eq:f2-}). Note
that, in contrast to the $j=1/2$ projection, this correction cannot
be expressed via elementary functions. This circumstance complicates
the derivation of the $j=3/2$ projection of the correction due to
the external magnetic field in $ns$ states with arbitrary $n$.

\subsection{Magnetic polarizability and nuclear magnetic moment shielding in the ground state of the hydrogen-like atom}

As an application of the obtained correction to the wave function,
let us first calculate the magnetic polarizability of the
hydrogen-like atom, which is related to a correction to the energy
level quadratic in a magnetic field.

Though the Hamiltonian (\ref{eq:ham_mag}) is linear in the magnetic
field, its eigenvalues are not. One can consider them at the
low-field limit (i.e. the field weak in comparison to the hyperfine
splitting), which contains terms non-linear in $\mathbf{B}$ (see,
e.g. \cite{bethe}).

Here we derive the polarizability for another case, when the
hyperfine splitting can be neglected (that is a kind of `medium'
field, which is weak comparing to the gross structure splittings,
but strong in respect to the nuclear hyperfine field). Once we
neglect the hyperfine splitting, the magnetic polarizability $\beta$
can be presented in the following form:
\begin{align*}
\delta E=\frac{\beta\, \mathbf{B}^2}{2} & = \left(
\frac{eB}{2}\right)^{2} \int d^3r\psi_{+}^{\dagger}\left(
\boldsymbol{\alpha}\times \boldsymbol{r}\right)_{z} G_{r}\left(
\boldsymbol{\alpha}\times
\boldsymbol{r}\right)_{z}\psi_{+}\\
& = \left( \frac{eB}{2}\right) \int d^3r\psi_{+}^{\dagger}
\left( \boldsymbol{\alpha}\times\boldsymbol{r}\right)_{z}\delta\psi_{B}%
\end{align*}

Using Eqs.~(\ref{eq:dpsi1(1/2)})--(\ref{eq:dpsi2(3/2)}), we obtain
\[
\beta=\frac{\alpha(Z\alpha)}{9m^2}\;C^{2}\int d^3r
\left( mr\right)^{2\gamma-1} \exp(-2Z\alpha mr)\left( g_{2+}-3g_{1+}\right) .
\]

Taking the integrals, we have
\begin{equation}\label{eq:kns1s}
\beta=\frac{\alpha\left( 2\gamma+1\right) }{36m^{3}\left( 1-\gamma\right)
}\left[ \frac{\left( \nu+\gamma\right) \left( 2\gamma+3\right) \left(
\nu+\gamma+1\right) }{3\left( \nu+\gamma+2\right) \left( \nu
+\gamma+3\right) } {}_{3}F_{2}\left( 2,2,\nu-\gamma;\nu+\gamma+4,\nu
-\gamma+1;1\right) +4\gamma-2\right]
\end{equation}
This result agrees with \cite{magpolcir}.

In the limit $Z\alpha \ll1$ we have
\[
\beta_{\mathrm{nr}}=\frac{\alpha}{m^{3}}\left( \frac{1}{2(Z\alpha )^{2}%
}-\frac{2}{3}+\ldots\right)\;.
\]

Let us now calculate the shielding of the nuclear magnetic moment in
the ground state. The correction to the energy of the ground state
in the external constant magnetic field reads
\begin{align*}
\delta E & =-\mu_{\mathrm{nucl}}Bk=\alpha\mu_{\mathrm{nucl}}B\int
d^3r\,\psi_{+}^{\dagger}\frac{\left( \boldsymbol{\alpha}%
\times\boldsymbol{r}\right)_{z}}{r^{3}}G_{r}\left( \boldsymbol{\alpha
}\times\boldsymbol{r}\right)_{z}\psi_{+}\,,\\
k & =\frac{2\alpha(Z\alpha)m}{9}C^{2}\int d^3r\,\left( mr\right)
^{2\gamma-4}\exp\left[ -2Z\alpha mr\right] \left( g_{2+}-3g_{1+}\right)\,,
\end{align*}
where we introduce the shielding constant $k$ as
\[
g^\prime=(1+k)g^{(0)}\,.
\]

Again, taking the integral, we obtain
\begin{equation}
\label{eq:k1s}k=\frac{4\alpha Z\alpha }{9}\left( \frac{1}{3}-\frac
{1}{6(1+\gamma )}+\frac{2}{\gamma }-\frac{3}{2\gamma -1}\right)\,,
\end{equation}
in agreement with \cite{moo,pyp,cjp02}.

\section{Shielding of the nuclear magnetic moment in $ns_{1/2}$ states}
\label{sec:iii}

Let us now calculate the corrections to the wave functions of the
$ns_{1/2}$ state for an arbitrary $n$. Using the projection operators
(\ref{eq:Projectors}), one can represent the correction to the
energy in the external constant magnetic field as
\begin{align}
\delta E & =\mu_{\mathrm{nucl}}B\,e^{2}\int d^3r\,\psi_{+}^{\dagger
}\frac{\left( \boldsymbol{\alpha}\times\boldsymbol{r}\right)_{z}}{r^{3}%
}G_{\mathrm{\mathrm{r}}}\left( \boldsymbol{\alpha}\times\boldsymbol{r}%
\right)_{z}\psi_{+}\nonumber\\
& =\mu_{\mathrm{nucl}}B\,e^{2}\int d^3r\,\psi_{+}^{\dagger}%
\frac{\left( \boldsymbol{\alpha}\times\boldsymbol{r}\right)_{z}}{r^{3}%
}G_{\mathrm{\mathrm{r}}}\left( P_{3/2}+P_{1/2}\right) \left(
\boldsymbol{\alpha}\times\boldsymbol{r}\right)_{z}\psi_{+}\nonumber\\
& =2\mu_{\mathrm{nucl}}\,e\int d^3r\,\psi_{+}^{\dagger}\frac{\left(
\boldsymbol{\alpha}\times\boldsymbol{r}\right)_{z}}{r^{3}}\delta\psi
_{B}^{1/2}+B\,e\int d^3r\,\left( \delta\psi_{\mu}^{3/2}\right)
^{\dagger}\left( \boldsymbol{\alpha}\times\boldsymbol{r}\right)_{z}\psi
_{+}\label{eq:deltaEns}\,.%
\end{align}

Here $\delta\psi_{\mu}$ and $\delta\psi_{B}$ denote the first-order
corrections to the wave function due to the magnetic moment of the
nucleus and due to the external field, respectively. As we shall
see, this representation is convenient for the calculations, since
both $\delta\psi_{\mu}^{3/2}$ and $\delta\psi_{B}^{1/2}$ are
expressed via the action of some operators, polynomial in $r$, on
the wave function $\psi_{ns_{1/2}}$ of the $ns_{1/2}$ state. This
wave function has the form
\begin{equation}
\psi_{ns_{1/2}}=C\,(mr)^{\gamma-1}\exp\left( -\lambda r\right) \left(
\begin{array}
[c]{c}%
a\sqrt{m+\varepsilon }\varphi_{+}\\
i b\sqrt{m-\varepsilon }\left( \boldsymbol{\sigma}\cdot\boldsymbol{n}\right)
\varphi_{+}%
\end{array}
\right) , \label{eq:psi_ns}%
\end{equation}
where $C$ is a normalization constant and the polynomials $a$ and
$b$ satisfy the following system of equations
\begin{align}
a^{\prime} & =\left( -\lambda -\frac{Z\alpha}{\lambda r}
\left(m-\varepsilon \right) \right) b+\left( \lambda+\frac{1-\gamma }%
{r}\right)a\,,\label{eq:ab}\\
b^{\prime} & =\left( -\lambda+\frac{Z\alpha}{\lambda r}\left(
m+\varepsilon \right) \right) a+\left( \lambda -\frac{\gamma +1}%
{r}\right) b\,,\nonumber\\
a_{-}^{\prime} & =\left( \frac{1}{r}-\frac{Z\alpha m}{\lambda r}\right)
a_{+}-\left( \frac{\gamma }{r}+\frac{Z\alpha\varepsilon}{\lambda r}
 -2\lambda \right) a_{-}\,,\nonumber\\
a_{+}^{\prime} & =\left( -\frac{\gamma}{r} +\frac{Z\alpha\varepsilon}{\lambda r} \right) a_{+}
+\left( \frac{1}{r}+\frac{Z\alpha m}{\lambda r}\right) a_{-}\,.\nonumber
\end{align}

Here $\varepsilon $ is the energy of the state, $\lambda
=\sqrt{m^{2} -\varepsilon^{2}}$, $a_{\pm}=a\pm b$. For $1s_{1/2}$
state, $a=b=1$. The idea of our derivation of the corrections to the
wave functions is to use these differential equations rather than
the explicit form of the functions $a$ and $b$.

\subsection{Calculation of $\delta\psi_{B}^{1/2}$}

We start from the derivation of the correction to the wave function
$\delta\psi_{B}^{1/2}$ due to the presence of the magnetic field.
Again, using the projectors (\ref{eq:Projectors}), we obtain
\[
\delta\psi_{B}^{1/2}=-C\,\frac{eB}{3m}G_{r}\,(mr)^{\gamma}\exp\left( -Z\alpha
mr\right) \left(
\begin{array}
[c]{c}%
b\sqrt{m-\varepsilon }\,\varphi_{+}\\
a\sqrt{m+\varepsilon }\,i\left( \boldsymbol{\sigma}\cdot\boldsymbol{n}\right)
\varphi_{+}%
\end{array}
\right) =\frac{eB}{3}G_{r}\left( -i\boldsymbol{\gamma }\cdot\boldsymbol{r}%
\right) \psi_{ns_{1/2}}\,.%
\]

Using the following identities
\begin{align*}
-im\boldsymbol{\gamma }\cdot\boldsymbol{r} & =\frac{3}{2}%
+\boldsymbol{\Sigma}\cdot\boldsymbol{L}+\frac{i}{2}\left[ \boldsymbol{\alpha
}\cdot\boldsymbol{r},H_{0}\right]\,, \\
\boldsymbol{L}^{2}\psi_{ns_{1/2}} & =\left( 1-\gamma^{0}\right)
\psi_{ns_{1/2}}%
\end{align*}
and the fact that $G_{r}\psi_{ns_{1/2}}=0$, we arrive at
\[
\delta\psi_{B}^{1/2}=\frac{eB}{3m}\left( G_{r}\gamma^{0}\psi_{ns_{1/2}}%
+\frac{i}{2}\boldsymbol{\alpha}\cdot\boldsymbol{r}\psi_{ns_{1/2}}\right)\,.
\]

The first term in the braces is conveniently represented as
\[
G_{r}\gamma^{0}\psi_{ns_{1/2}}=\frac{\partial}{\partial m}\psi_{ns_{1/2}%
}=\frac{1}{m} \left( \frac{3}{2}+r\frac{\partial}{\partial r}\right)
\psi_{ns_{1/2}}\,.
\]

Using Eqs.~(\ref{eq:psi_ns}) and~(\ref{eq:ab}), we have
\begin{equation}
\delta\psi_{B}^{1/2}=\frac{eB}{3m}\left( \frac{1}{2m}+\frac{\gamma^{0}}%
{m}+\left(\frac{ \varepsilon}{m} +\frac{1}{2}+\frac{Z\alpha }{mr}\right)
i\boldsymbol{\alpha}\cdot\boldsymbol{r}+i\boldsymbol{\gamma }\cdot
\boldsymbol{r}\right) \psi_{ns_{1/2}} \label{eq:deltaPsi12}\,.%
\end{equation}

\subsection{Calculation of $\delta\psi_{\mu}^{3/2}$}

Let us find the correction to the wave function due to the magnetic
field induced by the nuclear magnetic moment.
\begin{align*}
\delta\psi_{\mu}^{3/2} & =C\,e\mu_{\mathrm{nucl}}\,G_{r}\,(mr)^{\gamma-4}%
\exp\left( -\lambda r\right) \left(
\begin{array}
[c]{c}%
b\sqrt{m-\varepsilon}\;\left[ z\left( \boldsymbol{\sigma}\cdot
\boldsymbol{n}\right) -r/3\right] \varphi_{+}\\
a\sqrt{m+\varepsilon}\;i\left[ z-\left( \boldsymbol{\sigma}\cdot
\boldsymbol{r}\right) /3\right] \varphi_{+}%
\end{array}
\right) \\
& =C\,e\mu_{\mathrm{nucl}}\,(mr)^{\gamma-2}\exp\left( -\lambda r\right)
\left(
\begin{array}
[c]{c}%
g\sqrt{m+\varepsilon}\;\left[ z\left( \boldsymbol{\sigma}\cdot
\boldsymbol{n}\right) -r/3\right] \varphi_{+}\\
h\sqrt{m-\varepsilon}\;i\left[ z-\left( \boldsymbol{\sigma}\cdot
\boldsymbol{r}\right) /3\right] \varphi_{+}%
\end{array}
\right)\,.
\end{align*}

Acting by the operator $\varepsilon-H_0$, we obtain the following
system
\begin{align*}
g^{\prime} & =\left( \lambda -\frac{\gamma +2}{r}\right) g-\left(
\lambda +\frac{Z\alpha m}{\lambda r}\left(m-\varepsilon \right)
\right) h+\frac{a}{r^2}\,,\\
h^{\prime} & =\left( -\lambda +\frac{Z\alpha m}{\lambda r}\left(
m+\varepsilon \right) \right) g-\left( \frac{\gamma -2}{r}-\lambda
 \right) h-\frac{b}{r^2}\,.
\end{align*}

Adding and subtracting these equations, we have for $g_{\pm}=g\pm h$
\begin{align*}
g_{+}^{\prime} & =\frac{1}{r}\left( \frac{Z\alpha m}{\lambda }-2\right)
g_{-}-\frac{1}{r}\left( \gamma -\frac{Z\alpha}{\lambda }\varepsilon
 \right) g_{+}+\frac{a_{-}}{r^2}\,,\\
g_{-}^{\prime} & =-\frac{1}{r}\left( \frac{Z\alpha m}{\lambda}+2\right)
g_{+}+\left( 2\lambda -\frac{\gamma }{r}-\frac{Z\alpha}{\lambda r}
\varepsilon \right) g_{-}+\frac{a_{+}}{r^2}%
\end{align*}
and, after eliminating of $g_{-}$,
\[
-rg_{+}^{\prime\prime}-\left( 1+2\gamma -2\lambda r\right) g_{+}^{\prime
}-\left( 2\lambda (n-1)-\frac{3}{r}\right) g_{+}=\frac{a_{-}+a_{+}}{r^2}\,.%
\]

Keeping in mind that $a_{\pm}$ are some polynomials, we may search
for the solution of this system in the form of power series. It is
obvious, that the number of terms in these power series, if finite,
grows with $n$. Instead, we search for the solution in the form
$g_{+}=a_{+}\sum_{k}C_{k}r^{k}+a_{-} \sum_{k}D_{k}r^{k}$ and compare
the coefficients at $r^{k}a_{\pm}$.

We arrive at the following system of recurrence relations
\begin{align*}
-\left( k^{2}+2k\left( n-1+\gamma \right) -3\right) C_{k}-2k\left(
1-\frac{Z\alpha m}{\lambda }\right) D_{k} & =\delta_{k,-1}-2\left(
k-1\right) \lambda C_{k-1}\,,\\
-\left( k^{2}-2k\left( n-1+\gamma \right) -3\right) D_{k}-2k\left(
1+\frac{Z\alpha m}{\lambda }\right) C_{k} & =\delta_{k,-1}+2k\lambda
 D_{k-1}\,.%
\end{align*}

For $k=-1$ we have a system
\begin{gather*}
2\left( n+\gamma \right) C_{-1}+2\left( 1-\frac{Z\alpha m}{\lambda
 }\right) D_{-1}=1\,,\\
2\left( 2-n-\gamma \right) D_{-1}+2\left( 1+\frac{Z\alpha m}{\lambda
 }\right) C_{-1}=1\,.
\end{gather*}
with the solution
\[
C_{-1}=-D_{-1}=\frac{m-\varepsilon } {2\lambda Z\alpha}\,.\\
\]
For $k=0$, we have
\begin{gather*}
C_{0}=\frac{2\lambda}{3} C_{-1}\,,\\
D_{0}=0\,.
\end{gather*}
and, obviously, the coefficients for higher $k$ are equal to zero.

Thus, we obtain
\begin{align*}
g_{+} & =\frac{m-\varepsilon}{2Z\alpha }\left[ \left( \frac{1}%
{\lambda r}+\frac{2}{3}\right) a_{+}-\frac{1}{\lambda r}a_{-}\right]\,,\\
g_{-} & =\frac{m-\varepsilon}{2Z\alpha }\left[ \frac{1}{\lambda r}%
a_{+}-\left( \frac{1}{\lambda r}-\frac{2}{3}\right) a_{-}\right]
\end{align*}
and
\begin{align*}
g & =\frac{m-\varepsilon}{Z\alpha }\left[ \frac{1}{\lambda r}b+\frac
{1}{3}a\right]\,,\\
h & =\frac{m-\varepsilon }{3Z\alpha }b\,.
\end{align*}

In matrix notations the result can be presented as
\begin{equation}
\delta\psi_{\mu}^{3/2}=e\mu_{\mathrm{nucl}}\frac{m}{3Z\alpha }\left[
\frac{z}{r}\boldsymbol{\Sigma}\cdot\boldsymbol{n}-\frac{1}{3}\right]
\left[ 1-\frac{\varepsilon}{m} -\frac{3}{2mr}\left(
1+\gamma_{0}\right) i\boldsymbol{\gamma }\cdot\boldsymbol{n}\right]
\psi\label{eq:deltaPsi32}\;.
\end{equation}
For the particular case $n=2$ the expression agrees with
Eqs.~\ref{xd2s} and~\ref{yd2s} derived independently.

\subsection{Total result}

Using Eqs.~(\ref{eq:deltaEns}), (\ref{eq:deltaPsi12})
and~(\ref{eq:deltaPsi32}), we obtain the following result for the
shielding coefficient:
\[
k=-\frac{2\alpha}{9}\left\langle ns_{1/2}\right\vert \frac{1}{mr}\gamma
^{0}-2i\frac{\boldsymbol{\gamma }\cdot\boldsymbol{r}}{m^{2}r^{3}}%
-\frac{m-\varepsilon }{3Z\alpha }i\boldsymbol{\gamma }\cdot\boldsymbol{r}%
+\frac{1-\gamma^{0}}{2Z\alpha }\left\vert ns_{1/2}\right\rangle\,.
\]

The first two terms correspond to the contribution of $j=1/2$ while
the last two terms to that of $j=3/2$. This matrix element can be
evaluated using the recurrence relations given in
\cite{Shabaev1991}. The final result reads
\begin{align}
k & =-\frac{2\alpha}{9}\left[ \frac{2}{Z\alpha} \left( \frac{\lambda}{m} \right)^2
+ \frac
{4\left( 2\varepsilon +m\right) }{m\gamma \left( 4\gamma
^{2}-1\right) } \left( \frac{\lambda}{m} \right)^3 +\frac{\left( m-\varepsilon \right) \left(
m+2\varepsilon \right) }{6Z\alpha m^2 }+\frac{m-\varepsilon }{2Z\alpha m
}\right] \nonumber\label{eq:kns}\\
& =-\frac{2\alpha}{9}\left[ \left( \frac{\lambda}{m} \right)^3 \frac{ 4\left( 2\varepsilon
 +m\right) }{m\gamma \left( 4\gamma^{2}-1\right) }+\frac{\left(
m-\varepsilon \right) \left( 5m+4\varepsilon \right) }{3Z\alpha m^2 }\right]\,,
\end{align}
and its expansion for small $Z\alpha$ is
\[
k=-\frac{\alpha(Z\alpha)}{3n^{2}}\left[ 1+\frac{132n-35}{36n^{2}}(Z\alpha
)^{2}+\frac{370n^{3}+342n^{2}-492n+69}{72n^{4}}(Z\alpha)^{4}\right.
\]%
\begin{equation}
\left. +\frac{13208n^{5}+15048n^{4}-8552n^{3}-25320n^{2}+17064n-1641}%
{1728n^{6}}(Z\alpha)^{6}+\mathcal{O}\left( (Z\alpha)^{8}\right)
\right]\;.
\end{equation}
Only the first term of this expansion has been known for arbitrary
$n$, while the other terms were known only for the $1s$ state.

Note that the found correction corresponds to the following term in
the effective Hamiltonian
\[
\delta H=-k\left( \boldsymbol{B\mu}_{\mathrm{nucl}}\right)\,.
\]
Another structure, linear in the nuclear magnetic moment and magnetic field
\[
\left( \boldsymbol{B}\left[ \boldsymbol{\mu}_{\mathrm{nucl}}\times
\boldsymbol{S}_{\mathrm{e}}\right] \right)
\]
is $T$-odd and does not appear, which is consistent with the results
\cite{jetp01} (cf. also \cite{moo,pyp}).

\section*{Acknowledgments}

The work was supported in part by RFBR (under grant \# 06-02-04018) and by DFG (under the grant \# GZ 436 RUS 113/769/0-2). R.N.L.
thanks the  Max-Planck-Institut  f\"ur  Quantenoptik, Garching, for their hospitality and support.

\appendix

\section*{Appendices}

\section{Direct calculation of $\delta\psi^{1/2}_{\mu}$ for $2s$ state}

\label{app:b}

The wave function correction due to the nuclear magnetic moment can
be calculated using the method developed in \cite{Shabaev1991}. For
the $j=1/2$ part of the correction to the ground state it was
obtained in \cite{jetp00} confirming the result of \cite{pyp}. For
the $2s$ state one can obtain\footnote{Note a different
representation of the wave function correction as compared with the
above text. Here we follow \cite{jetp00,jetp01}, where the related
results were obtained for the $1s$ state in a similar manner.}
\begin{equation}
\delta\psi^{1/2}_{\mu}= \left(
\begin{array}
[c]{c}%
X^{(s)}_{2s}(r) \; \Omega_{1/2,0,m}(\boldsymbol{n})\\
- i (\boldsymbol{\sigma}\cdot\boldsymbol{n}) \, Y^{(s)}_{2s}(r) \; \Omega
_{1/2,0,m}(\boldsymbol{n})
\end{array}
\right) \,,
\end{equation}
where
\begin{equation}
\psi_{2s}= \left(
\begin{array}
[c]{c}%
f_{2s}(r) \; \Omega_{1/2,0,j_z}(\boldsymbol{n})\\
- i (\boldsymbol{\sigma}\cdot\boldsymbol{n}) \, g_{2s}(r) \; \Omega
_{1/2,0,j_z}(\boldsymbol{n})
\end{array}
\right)
\end{equation}
is the unperturbed state, $\Omega_{j,l,m}(\boldsymbol{n})$ are the
spherical spinors,
\begin{align}
X^{(s)}_{2s} & = \frac{1}{1-4\gamma^{2}} \left[ \left( \frac{2\lambda^{3}%
}{m^{2}\gamma}-\frac{3}{r} \right) f_{2s} + \left( -3m(\varepsilon/m +1) -
\frac{2Z\alpha}{r} \right) g_{2s} \right. \nonumber\\
& - \left. \frac{2Z\alpha m(2\varepsilon/m +1)}{r}F \right] \,,\\
Y^{(s)}_{2s} & = \frac{1}{1-4\gamma^{2}} \left[ \left( \frac{2\lambda^{3}%
}{m^{2}\gamma}+\frac{3}{r} \right) g_{2s} + \left( 3m(\varepsilon/m -1) +
\frac{6Z\alpha}{r} \right) f_{2s} \right. \nonumber\\
& - \left. \frac{2Z\alpha m(2\varepsilon/m +1)}{r}G \right] \,,
\end{align}
\begin{align}
F & = N (2\lambda r)^{\gamma} e^{-\lambda r} \sqrt{2\lambda
/m+Z\alpha} \left[ \frac{(Z\alpha m)^{4}-2(Z\alpha m)^{3} \lambda+ Z\alpha m
\lambda^{3} - 2\lambda^{4}} {2(Z\alpha m-\lambda) \lambda^{3}} +
\frac{2Z\alpha m\;\psi(1+2\gamma)}{\lambda} \right. \nonumber\\
& \left. + \left( \frac{-2(Z\alpha m)^{4} + 4(Z\alpha m)^{3}\lambda- 5
(Z\alpha m)^{2}\lambda^{2} + 3Z\alpha m\lambda^{3} - 2\lambda^{4}} {Z\alpha
m\lambda(Z\alpha m-\lambda)^{2}} - \frac{4\lambda\;\psi(1+2\gamma)}{Z\alpha
m-\lambda} \right) \lambda r \right. \nonumber\\
& \left. + \frac{2\lambda}{Z\alpha m-\lambda} \; (\lambda r)^{2} -
\frac{2Z\alpha m}{\lambda} \; \log\bigl( 2 \lambda r ) + \frac{4\lambda
}{Z\alpha m-\lambda} \; \lambda r \log\bigl( 2 \lambda r \bigr) \right] \,,\\
G & = - N (2\lambda r)^{\gamma} e^{-\lambda r} \sqrt
{2\lambda/m-Z\alpha} \left[ \frac{(Z\alpha m)^{5}+5(Z\alpha m)^{2}
\lambda^{3} - 8Z\alpha m \lambda^{4} - 4\lambda^{5}} {2Z\alpha m(Z\alpha
m-\lambda) \lambda^{3}} \right. \nonumber\\
& + \frac{2(Z\alpha m+2\lambda)\;\psi(1+2\gamma)}{\lambda}\nonumber\\
& \left. + \left( \frac{-2(Z\alpha m)^{4} + 2(Z\alpha m)^{3}\lambda-
(Z\alpha m)^{2}\lambda^{2} - 3Z\alpha m\lambda^{3} + 2\lambda^{4}} {Z\alpha m
\lambda(Z\alpha m-\lambda)^{2}} - \frac{4\lambda\;\psi(1+2\gamma)}{Z\alpha
m-\lambda} \right) \lambda r \right. \nonumber\\
& \left. + \frac{2\lambda}{Z\alpha m-\lambda} \; (\lambda r)^{2} -
\frac{2(Z\alpha m+2\lambda)}{\lambda} \; \log\bigl( 2 \lambda r ) +
\frac{4\lambda}{Z\alpha m-\lambda} \; \lambda r \log\bigl( 2 \lambda r
\bigr) \right] \;,
\end{align}
\[
N = \frac{\lambda^{3}}{2((Z\alpha m)^{2}-2\lambda^{2})} \; \sqrt
{\frac{1+2\gamma}{Z\alpha m(Z\alpha m+\lambda)\Gamma(1+2\gamma)}} \,,
\]
\[
\psi(z)=\Gamma^{\prime}(z)/\Gamma(z) \,.
\]

The result for the $j=1/2$ part of the $2s$ shielding obtained using
this expression agrees with Eq.~(\ref{eq:kns}).

The same method can be used for the calculation of the $j=3/2$ part
of the correction for any $ns$ state and it yields
\begin{equation}
\delta\psi^{3/2}_{\mu}= \left(
\begin{array}
[c]{c}%
X^{(d)}_{ns}(r) \; \Omega_{3/2,2,m}(\boldsymbol{n})\\
- i (\boldsymbol{\sigma}\cdot\boldsymbol{n}) \, Y^{(d)}_{ns}(r) \; \Omega
_{3/2,2,m}(\boldsymbol{n})
\end{array}
\right) \,,
\end{equation}
where
\begin{align}
X^{(d)}_{ns} & =\frac{1}{Z\alpha}\left( \frac{\varepsilon -m}{3}f_{ns}
+\frac{g_{ns}}{r}\right) \,,\label{xd2s}\\
Y^{(d)}_{ns} & = \frac{1}{Z\alpha} \frac{\varepsilon -m}{3}g_{ns}\,,\label{yd2s}
\end{align}
which agrees with Eq.~(\ref{eq:deltaPsi32}).

\end{document}